\documentclass[twocolumn, tighten]{aastex7}

\usepackage{amsmath}

\shorttitle{Broadband F070W LyS Opacity}
\shortauthors{Zhu et al.}

\begin{document}

\title{Excess Broadband Lyman-Series Opacity in Deep JWST/NIRCam Imaging: Constraints on Galaxy-associated \ion{H}{1} Absorption}

\author[0000-0003-3307-7525]{Yongda Zhu}
\affiliation{Steward Observatory, University of Arizona, 933 North Cherry Avenue, Tucson, AZ 85721, USA}
\email[show]{yongdaz@arizona.edu}

\author[0000-0003-1344-9475]{Eiichi Egami}
\affiliation{Steward Observatory, University of Arizona, 933 North Cherry Avenue, Tucson, AZ 85721, USA}
\email{egami@arizona.edu}

\author[0000-0002-7893-6170]{Marcia J.\ Rieke}
\affiliation{Steward Observatory, University of Arizona, 933 North Cherry Avenue, Tucson, AZ 85721, USA}
\email{mrieke@arizona.edu}

\author[0000-0001-5211-1958]{Laura C. Keating}
\affiliation{Institute for Astronomy, University of Edinburgh, Blackford Hill, Edinburgh, EH9 3HJ, UK}
\email[]{Laura.Keating@ed.ac.uk}

\author[0000-0001-9420-7384]{Christopher Cain}
\affiliation{Institute of Astronomy, University of Cambridge, Madingley Road, Cambridge CB3 0HA, UK}
\affiliation{KICC - Kavli Institute for Cosmology Cambridge, Madingley Road, Cambridge CB3 0HA, UK}
\affiliation{School of Earth and Space Exploration, Arizona State University, Tempe, AZ 85287-6004, USA}
\email{clcain3@asu.edu}

\author[0000-0003-3310-0131]{Xiaohui Fan}
\affiliation{Steward Observatory, University of Arizona, 933 North Cherry Avenue, Tucson, AZ 85721, USA}
\email{xfan@arizona.edu}

\author[0000-0003-2344-263X]{George D. Becker}
\affiliation{Department of Physics \& Astronomy, University of California, Riverside, CA 92521, USA}
\email{georgeb@ucr.edu}

\author[0000-0002-8651-9879]{Andrew J.\ Bunker}
\affiliation{Department of Physics, University of Oxford, Denys Wilkinson Building, Keble Road, Oxford OX1 3RH, UK}
\email{andy.bunker@physics.ox.ac.uk}

\author[0000-0002-4622-6617]{Fengwu Sun}
\affiliation{Department of Astronomy, Westlake University, Hangzhou 310030, Zhejiang Province, China}
\affiliation{Center for Astrophysics $|$ Harvard \& Smithsonian, 60 Garden St., Cambridge, MA 02138, USA}
\email{sunfengwu@westlake.edu.cn}

\author{Anson D'Aloisio}
\affiliation{Department of Physics \& Astronomy, University of California, Riverside, CA 92521, USA}
\email{ansond@ucr.edu}

\author{Esteban Moreno}
\affiliation{School of Earth and Space Exploration, Arizona State University, Tempe, AZ 85287-6004, USA}
\email{emoren47@asu.edu}


\author[0000-0003-2764-8248]{James S. Bolton}
\affiliation{School of Physics and Astronomy, University of Nottingham, University Park, Nottingham, NG7 2RD, UK}
\email{james.bolton@nottingham.ac.uk}

\author[0000-0003-3693-3091]{Valentina D'Odorico}
\affiliation{INAF-Osservatorio Astronomico di Trieste, Via Tiepolo 11, I-34143 Trieste, Italy}
\affiliation{IFPU-Institute for Fundamental Physics of the Universe, via Beirut 2, I-34151 Trieste, Italy}
\email{valentina.dodorico@inaf.it}

\author[0000-0002-2929-3121]{Daniel J.\ Eisenstein}
\affiliation{Center for Astrophysics $|$ Harvard \& Smithsonian, 60 Garden St., Cambridge, MA 02138, USA}
\email{deisenstein@cfa.harvard.edu}

\author[0000-0001-7440-8832]{Yoshinobu Fudamoto}
\affiliation{Center for Frontier Science, Chiba University, 1-33 Yayoi-cho, Inage-ku, Chiba 263-8522, Japan}
\email{}

\author[0000-0001-8443-2393]{Martin G. Haehnelt}
\affiliation{Kavli Institute for Cosmology and Institute of Astronomy, Madingley Road, Cambridge, CB3 0HA, UK}
\email{haehnelt@ast.cam.ac.uk}

\author[0000-0003-4565-8239]{Kevin Hainline}
\affiliation{Steward Observatory, University of Arizona, 933 North Cherry Avenue, Tucson, AZ 85721, USA}
\email{kevinhainline@arizona.edu}

\author[0009-0008-7862-5277]{Seyedazim Hashemi}
\affiliation{Department of Physics \& Astronomy, University of California, Riverside, CA 92521, USA}
\email{seyedazim.hashemi@email.ucr.edu}

\author[0000-0001-7673-2257]{Zhiyuan Ji}
\affiliation{Steward Observatory, University of Arizona, 933 North Cherry Avenue, Tucson, AZ 85721, USA}
\email{zhiyuanji@arizona.edu}

\author[0000-0001-5829-4716]{Girish Kulkarni}
\affiliation{Tata Institute of Fundamental Research, Homi Bhabha Road, Mumbai 400005, India}
\email{kulkarni@theory.tifr.res.in}

\author[0000-0001-6052-4234]{Xiaojing Lin}
\affiliation{Department of Astronomy, Tsinghua University, Beijing 100084, China}
\email{}

\author[0009-0003-5402-4809]{Zheng Ma}
\affiliation{Steward Observatory, University of Arizona, 933 North Cherry Avenue, Tucson, AZ 85721, USA}
\email{mazh@arizona.edu}

\author[0000-0003-4924-5941]{Maria Pudoka}
\affiliation{Steward Observatory, University of Arizona, 933 North Cherry Avenue, Tucson, AZ 85721, USA}
\email{pudoka@arizona.edu}

\author[0000-0003-2303-6519]{George H.\ Rieke}
\affiliation{Steward Observatory, University of Arizona, 933 North Cherry Avenue, Tucson, AZ 85721, USA}
\email{grieke@arizona.edu}

\author[0000-0002-5104-8245]{Pierluigi Rinaldi}
\affiliation{Space Telescope Science Institute, 3700 San Martin Drive, Baltimore, Maryland 21218, USA}
\email{prinaldi@stsci.edu}

\author[0000-0002-4271-0364]{Brant Robertson}
\affiliation{Department of Astronomy and Astrophysics, University of California, Santa Cruz, 1156 High Street, Santa Cruz, CA 95064, USA}
\email{brant@ucsc.edu}

\author[0000-0001-9262-9997]{Christopher N.\ A.\ Willmer}
\affiliation{Steward Observatory, University of Arizona, 933 North Cherry Avenue, Tucson, AZ 85721, USA}
\email{cnaw@arizona.edu}

\author[0000-0003-0111-8249]{Yunjing Wu}
\affiliation{Kavli Institute for the Physics and Mathematics of the Universe (WPI), The University of Tokyo Institutes for Advanced Study, The University of Tokyo, Kashiwa, Chiba 277-8583, Japan}
\affiliation{Center for Data-Driven Discovery, Kavli IPMU (WPI), UTIAS, The University of Tokyo, Kashiwa, Chiba 277-8583, Japan}
\email{yunjing.wu@ipmu.jp}

\author[0000-0002-1574-2045]{Junyu Zhang}
\affiliation{Steward Observatory, University of Arizona, 933 North Cherry Avenue, Tucson, AZ 85721, USA}
\email{junyuzhang@arizona.edu}

\begin{abstract}
The rapid saturation of the Ly$\alpha$ forest limits our view of the end of cosmic reionization. We measure broadband Lyman-series (LyS) transmission toward 279 galaxies at $5.45<z<6.85$ using deep JWST/NIRCam F070W imaging from JADES. F070W combines foreground Ly$\alpha$ with Ly$\beta$, Ly$\gamma$, and higher-order absorption along each galaxy sightline. The mean transmission declines from $0.120$ at $z=5.55$ to $0.013$ at $z=6.55$. Assuming a power-law intrinsic continuum, the late-reionization Sherwood-Relics simulation reproduces this decline better than a fully ionized IGM with a homogeneous ultraviolet background, but still requires about 8\% additional opacity. A possible interpretation of this additional opacity is the contribution to the optical depth from Damped-Ly$\alpha$ absorbers (DLAs) hosted within the source galaxies. By fitting a simple model which includes this contribution combined with the mean LyS flux of a late-ending reionization simulation, we constrain the mean H\,\textsc{i} column density of a galaxy DLA to be $\log_{10}(N_{\rm HI}/{\rm cm}^{-2})=22.3^{+0.2}_{-0.3}$. Our results extend Lyman-series transmission measurements to hundreds of galaxy sightlines and provide a new method for characterising the neutral gas content of galaxies complementary to spectroscopic studies.
\end{abstract}


\section{Introduction} \label{sec:intro}

The intergalactic medium (IGM) contains most of the baryons in the universe and forms the reservoir from which galaxies assemble and evolve. The Lyman-series (LyS) forest is a sensitive record of the ionization state of the IGM, providing a direct view of neutral hydrogen along the line of sight \citep[e.g.,][]{becker_reionisation_2015,mcquinn_evolution_2016,fan_quasars_2023,fan_constraining_2006,becker_evidence_2015,bosman_hydrogen_2022,yang_measurements_2020,zhu_chasing_2021,zhu_long_2022}. The transition from a mostly neutral to a mostly ionized IGM, known as reionization, is a key phase in cosmic history, with implications for galaxy formation, the growth of large-scale structure, and cosmological parameters inferred from the cosmic microwave background (CMB).
 
The presence of transmitted Ly$\alpha$ spikes near $z\sim6$ has long been interpreted as evidence that reionization was largely complete by this epoch. Yet the ionizing mean free path remained short \citep{becker_mean_2021,zhu_probing_2023,gaikwad_measuring_2023,davies_constraints_2024}, favoring a late and rapid end to reionization \citep{cain_short_2021,lewis_short_2022,davies_predicament_2021}. Maintaining a highly ionized IGM required continued ionizing emission from galaxies in the presence of optically thick absorbers, which regulate the mean free path and can drive ultraviolet-background (UVB) fluctuations.

The Thomson optical depth measured from the cosmic microwave background, $\tau_{\rm CMB}$, reflects the integrated ionization history. Planck finds $\tau_{\rm CMB}\approx0.054$--0.058 \citep{planck_collaboration_planck_2020,tristram_cosmological_2024}. However, larger values, $\tau_{\rm CMB}\sim0.09$, have been considered in connection with recent DESI BAO and CMB cosmological constraints \citep{sailer_addressing_2026,jhaveri_raising_2026}; see also \citet{cain_chasing_2025}, \citet{cain_cosmic_2025}, \citet{asthana_late-end_2024}, \citet{elbers_rapid_2025}, and \citet{garcia-gallego_constraints_2026}. Observations at $z\simeq5$--6 help anchor the end of reionization and constrain the ionization history at earlier times \citep{cain_chasing_2025}.

Several observations suggest that the transition to a transparent IGM extended below $z=6$. The Ly$\alpha$ forest shows large sightline-to-sightline fluctuations in effective optical depth and long dark troughs extending to $z\simeq5.5$ \citep[e.g.,][]{fan_constraining_2006,becker_evidence_2015,eilers_opacity_2018,bosman_new_2018,bosman_hydrogen_2022,yang_measurements_2020,zhu_chasing_2021,zhu_long_2022,wolfson_measurements_2024}. The ionizing mean free path evolves rapidly over the same redshift range \citep{worseck_giant_2014,becker_mean_2021,zhu_probing_2023,gaikwad_measuring_2023}. Galaxy surveys around opaque and transmissive quasar sightlines also point to environmental fluctuations in Ly$\alpha$ transmission \citep[e.g.,][]{becker_evidence_2018,kashino_evidence_2020,christenson_constraints_2021, ishimoto_physical_2022,christenson_relationship_2023,meyer_role_2020,kashino_eiger_2025,kakiichi_jwst_2025,zhu_galaxy_2026}. These results motivate models with a late end of reionization, fluctuations in the ultraviolet background, and/or a persistent population of optically thick absorbers \citep{daloisio_large_2015,davies_large_2016,chardin_large-scale_2017,daloisio_large_2018,davies_determining_2018,nasir_observing_2020,kulkarni_large_2019,keating_long_2020,choudhury_studying_2021,qin_reionization_2021,gnedin_cosmic_2022-1,cain_morphology_2023}.

The main observational challenge is that Ly$\alpha$ saturates quickly, even when the neutral hydrogen fraction is as low as $x_{\rm HI}\sim10^{-3}$. Once the transmitted Ly$\alpha$ flux approaches zero, different neutral-gas conditions cannot be distinguished from Ly$\alpha$ transmission alone. Higher-order Lyman-series absorption is less saturated: Ly$\beta$ and higher transitions have lower optical depths and remain sensitive to additional neutral opacity after Ly$\alpha$ loses discriminatory power. Ly$\beta$ dark gaps and Ly$\beta$ effective-opacity measurements show that high-opacity regions persist below $z=6$ \citep[e.g.,][]{mcgreer_model-independent_2015,barnett_observations_2017,eilers_anomaly_2019,yang_measurements_2020,zhu_long_2022,jin_nearly_2023,davies_updated_2025}, which is also seen in late-ending reionisation simulations \citep{keating_constraining_2020}.

Galaxies can also serve as background sources for IGM absorption, both spectroscopically and photometrically \citep[e.g.,][]{thomas_less_2021,kakiichi_photometric_2023,umeda_probing_2026,meyer_probing_2025,hu_using_2026}. Deep JWST/NIRCam imaging extends higher-order LyS measurements to many galaxy sightlines. For galaxies at $z\simeq5.5$--6.8, F070W probes rest-frame wavelengths largely below Ly$\alpha$, sampling a filter-weighted combination of foreground Ly$\alpha$, Ly$\beta$, Ly$\gamma$, Ly$\delta$, and higher-order absorption \citep{madau_radiative_1995}. JADES \citep{rieke_jades_2023,eisenstein_overview_2026,johnson_jwst_2026,robertson_jwst_2026} provides a sample deep enough to measure the mean F070W transmission to a precision of $\sim0.01$ per $\Delta z=0.2$ redshift bin, while SAPPHIRES \citep[Early Data Release, EDR;][]{sun_slitless_2025} provides an independent comparison.

We use these measurements to examine the IGM opacity near the end of reionization and the possible contribution from neutral gas associated with the galaxies. We compare the measurements with the fully ionized Sherwood simulation \citep{bolton_sherwood_2017}, the Sherwood-Relics late-reionization simulations \citep{puchwein_sherwood-relics_2023}, and three FlexRT histories \citep{cain_chasing_2025,gangolli_correlation_2025,cain_flexrt_2024}. Separating higher-order absorption from foreground Ly$\alpha$ in this comparison depends on the assumed Ly$\alpha$ transmission model. Galaxy sightlines also pass through H\,\textsc{i} in the interstellar and circumgalactic media (ISM and CGM) of the source galaxies themselves, which can host damped Ly$\alpha$ absorbers \citep[DLAs; e.g.,][]{heintz_extreme_2023,umeda_probing_2026,mason_constraints_2026}. Constraining this source-associated absorption matters for damping-wing measurements of the reionization history, and the broadband LyS opacity provides a statistical constraint on its mean column density.

Section~\ref{sec:data} describes the data and models, Section~\ref{sec:results} presents the transmission measurements, Section~\ref{sec:discussion} discusses their interpretation and constrains source-associated H\,\textsc{i} absorption, and Section~\ref{sec:conclusions} summarizes our conclusions. Appendix~\ref{app:model_order_transmission} presents the model Lyman-series transmission grids, Appendix~\ref{app:footprints} the survey footprints, Appendix~\ref{app:photoz_robustness} the effect of photometric-redshift errors, Appendix~\ref{app:aperture_robustness} the aperture dependence, and Appendix~\ref{app:continuum_robustness} tests of the intrinsic continuum. We assume a flat $\Lambda$CDM cosmology with $\Omega_m=0.3$, $\Omega_\Lambda=0.7$, and $H_0=70\ {\rm km\ s^{-1}\ Mpc^{-1}}$.

\section{Data and Models} \label{sec:data}

\begin{figure*}[!ht]
    \centering
    \includegraphics[width=\textwidth]{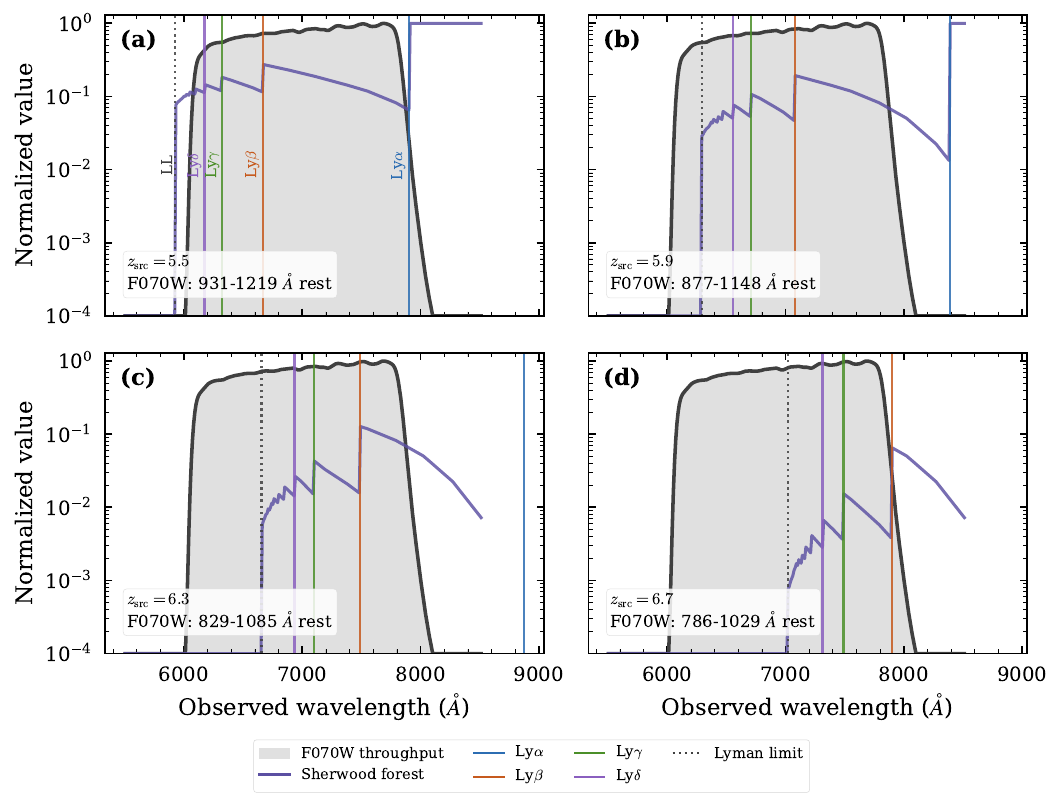}
    \caption{Wavelength coverage of the F070W LyS measurement. Panels (a)--(d) show the normalized F070W throughput for representative source redshifts, along with the expected Sherwood forest transmission and the observed-frame positions of Ly$\alpha$, Ly$\beta$, Ly$\gamma$, Ly$\delta$, and the Lyman limit. F070W samples a changing mixture of foreground Lyman-series absorption as source redshift increases.
}
    \label{fig:filter_geometry}
\end{figure*}

\begin{figure*}[!ht]
    \centering
    \includegraphics[width=\textwidth]{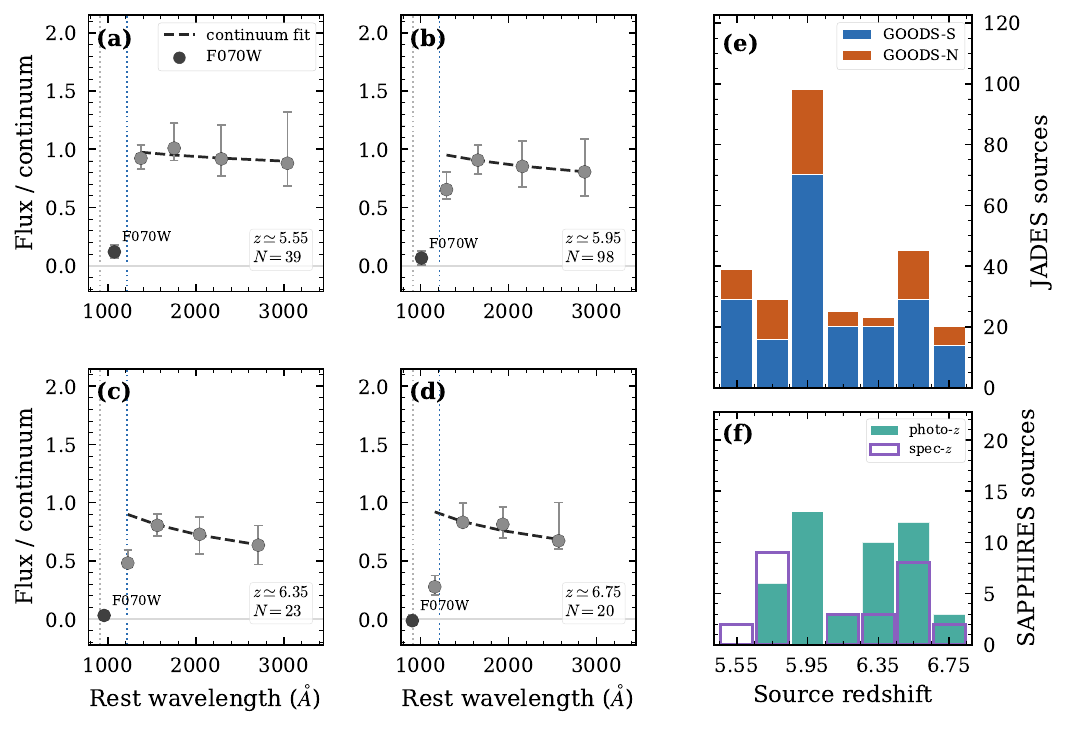}
   \caption{
    Photometric stacks and source counts for the F070W measurement. Panels (a)--(d) show illustrative median stacks of the selected JADES sources in four redshift bins, with each source normalized by its power-law continuum fitted at the band pivot wavelengths and evaluated at F070W; error bars show the 16th--84th percentile range of the sources. The dashed line shows the continuum fit to the redward bands, and the F070W point shows the LyS-suppressed flux. F090W, which straddles Ly$\alpha$ for $z_a>5.54$, is shown but excluded from the fit. Dotted vertical lines mark the Lyman limit and Ly$\alpha$. The binned transmissions in Table~\ref{tab:jades_binned} use the full-band continuum and the clipped bootstrap mean instead. Panels (e) and (f) show the number of selected sources per redshift bin: GOODS-S and GOODS-N stacked in panel (e), and the SAPPHIRES photo-$z$ (filled) and spec-$z$ (open) samples in panel (f). Each bar spans its $\Delta z=0.2$ bin.
}
    \label{fig:photometric_stacks}
\end{figure*}

Figure~\ref{fig:filter_geometry} shows the F070W wavelength coverage, and Figure~\ref{fig:photometric_stacks} shows the photometric stacks and source counts. The JADES and SAPPHIRES footprints are shown in Appendix~\ref{app:footprints}.

\subsection{JADES GOODS-S and GOODS-N Photometry} \label{sec:data_jades}

Our galaxy sample is drawn from the JADES DR5 GOODS-S and GOODS-N catalogs, which provide photometry and photometric-redshift information \citep{johnson_jwst_2026,robertson_jwst_2026,eisenstein_overview_2026,hainline_jwst_2026}. We use the catalog redshift $z_a$ and require finite $\chi_{\rm peak}<50$, where $\chi_{\rm peak}$ is the minimum template-fit $\chi^2$ of the catalog photometric-redshift solution, as a fit-quality criterion.

For the fiducial measurement we use CIRC1 aperture fluxes (aperture radius $r=0.1$ arcsec) measured on the mosaics PSF-matched to F444W, after local background subtraction and aperture correction to point-source total flux with the F444W PSF \citep{robertson_jwst_2026}. The same aperture is used for F070W and every continuum band, giving a consistent color measurement across the observed and continuum filters.

We select galaxies at $5.45<z_a<6.85$ and bin them from $z=5.55$ to 6.75 in intervals of $\Delta z=0.2$. The eligible redward bands are F090W, F115W, F150W, and F200W when their full filter window lies redward of source Ly$\alpha$. F090W is eligible only for $z_a<5.54$ under this criterion and is otherwise excluded. We require at least two positive eligible measurements and fit a power law, $f_\nu\propto\lambda^{\alpha}$, to their band-averaged fluxes; $F_{\rm F070W,cont}$ is this continuum averaged over the full F070W response. With the observed CIRC1 flux $F_{\rm F070W,obs}$, we define $T_{\rm F070W}=F_{\rm F070W,obs}/F_{\rm F070W,cont}$. We further require $10\leq F_{\rm F200W}\leq 1000$~nJy, $25\leq F_{\rm F070W,cont}\leq 1000$~nJy, and a valid F070W measurement (a finite F070W uncertainty, which removes sources without F070W coverage in the PSF-matched mosaic). The final sample contains 279 galaxies: 198 in GOODS-S and 81 in GOODS-N, with redshift-bin counts of 39, 29, 98, 25, 23, 45, and 20. GOODS-S and GOODS-N are analyzed separately and then combined to expose field-to-field differences. No selection is made using measured $T_{\rm F070W}$.

\subsection{SAPPHIRES EDR Imaging and Spectroscopic Redshifts} \label{sec:data_sapphires}

We also use the SAPPHIRES Early Data Release as an external comparison sample \citep{sun_slitless_2025}. SAPPHIRES combines NIRCam imaging and wide-field slitless spectroscopy (WFSS) obtained in parallel to observations of the Frontier Field cluster MACS~J0416.1$-$2403. The observations were taken as pure-parallel data to the NIRSpec multi-object spectroscopy program JWST-GO-4750 (PI: K.~Nakajima). With a total dual-channel exposure time of 47.2 hr, SAPPHIRES provides deep imaging over $0.6$--$5.0\,\mu$m and WFSS coverage over $3.1$--$5.0\,\mu$m using the F356W and F444W filters with orthogonal grism dispersion directions.

The imaging covers a total area of 16.9 arcmin$^2$ in 13 NIRCam filters: F070W, F090W, F115W, F140M, F150W, F182M, F200W, and F210M in the short-wavelength channel, and F277W, F335M, F356W, F410M, and F444W in the long-wavelength channel. The point-source $5\sigma$ depths are approximately 28.4--29.5 AB mag across the filters. We use the released SAPPHIRES photometric catalog and the associated spectroscopic catalog, matching them by source identifier when possible and by sky position otherwise.

We analyze the SAPPHIRES photo-$z$ and spec-$z$ samples separately. Applying the same redshift range, redward-continuum criteria, and F070W-validity requirement, plus the SAPPHIRES photometric quality cuts, gives 47 photo-$z$ and 27 spec-$z$ sources for comparison.

\subsection{Quasar Composite Measurements} \label{sec:data_quasar}

We compare the imaging-based F070W transmission to high-redshift quasar composite spectra from \citet{zhu_probing_2023}. That work measured the ionizing photon mean free path over $5<z<6$ using 97 quasar spectra at $4.9<z<6.1$, combining Keck/LRIS \citep{becker_mean_2021}, Gemini/GMOS, Keck/ESI, VLT/X-Shooter, XQR-30 \citep{dodorico_xqr-30_2023}, and new Keck/ESI observations. The spectra were shifted to the rest frame, normalized by the continuum flux over $1270$--$1380$~\AA, and stacked in redshift bins of width $\Delta z=0.3$.

The full quasar analysis contains four composite bins with mean redshifts $\langle z\rangle=5.08$, 5.31, 5.65, and 5.93, containing 44, 26, 9, and 18 spectra, respectively. In this paper we use only the original two higher-redshift stacks, at $\langle z\rangle=5.65$ and 5.93. S.~Hashemi et al.\ (in preparation) instead use an expanded sample with additional quasar spectra. These bins most closely overlap the absorber-redshift range probed by F070W for our galaxy sample.

\subsection{Opacity Models} \label{sec:data_models}

We compare the measurements to reference opacity models. The first is a  simulation from the Sherwood simulation suite \citep{bolton_sherwood_2017}, which was performed with a uniform UVB \citep{haardt_radiative_2012} and is already fully ionized over the redshift range considered here. Sherwood was run with the smoothed-particle hydrodynamics code \textsc{p-gadget-3}, an updated version of \textsc{gadget-2} \citep{springel_cosmological_2005}. In the specific simulation analysed here, the simulation has $2\times2048^3$ gas and dark-matter particles in a $(160\,h^{-1}{\rm cMpc})^3$ box, where $h=H_0/(100\ {\rm km\ s^{-1}\ Mpc^{-1}})$. As reionization completes around redshift 10, by $z\simeq6$ the IGM has had time to relax hydrodynamically.

The primary late-reionization comparison uses the Sherwood-Relics simulations \citep{puchwein_sherwood-relics_2023}. Sherwood-Relics uses a hybrid approach to model spatial variations in the ionization and thermal history of the IGM and the hydrodynamic response of the gas to patchy reionization, by using outputs from a radiative transfer simulation performed with \textsc{aton} \citep{aubert_radiative_2008} to impose an inhomogeneous UVB in the simulation. We use the late-reionization history in which reionization completes at $z\simeq5.3$. For this work, mean transmissions for the first 30 Lyman-series transitions were computed from Sherwood-Relics on a fine grid of absorber redshifts, $z_{\rm abs}=4.1$--$7.1$, with $\Delta z=0.1$.

We also include three reionization histories modeled with the \textsc{FlexRT} radiative-transfer code \citep{cain_flexrt_2024,gangolli_correlation_2025}. These simulations follow the evolution of ionized regions and IGM temperatures in a $(200\,h^{-1}\,\mathrm{cMpc})^3$ volume, using a $200^3$ radiative-transfer grid and a sub-grid prescription for unresolved ionizing absorbers. The late-start and early-start histories of \citet{cain_chasing_2025} both complete reionization at $z\simeq5$, with midpoints at $z\simeq6.5$ and $z\simeq7.5$, respectively. The very-early-start history of \citet{gangolli_correlation_2025} begins reionization earlier but is likewise calibrated to the Ly$\alpha$ forest at $z<6$, with a global neutral fraction of 6\% at $z=5.7$. For each history, the first 30 Lyman-series transitions are computed along sightlines originating from simulated halos, including the damping-wing absorption of neutral gas. The Ly$\alpha$ opacity in ionized gas was rescaled to match the mean Ly$\alpha$ transmission of the late-reionization model of \citet{kulkarni_large_2019}, excluding cells with neutral fractions above 10\%. We use the sightline-averaged transmissions over $5.03\leq z_{\rm abs}\leq7.93$.

All models are expressed as mean transmissions per transition, $\langle F_j\rangle(z_{\rm abs})$, which we convert to effective optical depths, $\tau_{{\rm eff},j}=-\ln\langle F_j\rangle$, and interpolate linearly in redshift. Because F070W also samples foreground Ly$\alpha$ at $z_{\rm abs}<5$, where not all reionization models are available, every model uses the Sherwood grid below $z_{\rm abs}=5$ and its own grid above; replacing this switch with a linear blend over $\Delta z=0.4$ changes the F070W predictions by less than 3\%. In this common foreground grid, the Ly$\alpha$ optical depths are rescaled to reproduce the observed mean Ly$\alpha$ forest transmission \citep{becker_refined_2013,boera_revealing_2019}, and the higher-order Lyman-series transmissions computed from the same rescaled skewers remain sensitive to the simulated distribution of H\,\textsc{i} optical depths, including the treatment of dense gas (Appendix~\ref{app:model_order_transmission}). At $z_{\rm abs}\geq5$, the Sherwood Ly$\alpha$ optical depths are rescaled to the mean Ly$\alpha$ transmission of the \citet{kulkarni_large_2019} model, which is broadly consistent with the measurements of \citet{bosman_hydrogen_2022}. The Sherwood higher-order opacities are obtained by scaling the Ly$\alpha$ optical depths with $f_j\lambda_j$, which does not reproduce the damping wings of neutral gas \citep[e.g.,][]{malloy_how_2015}. The FlexRT sightlines start at halos and therefore include the ionized environments of the host galaxies, which can change the transmission relative to the volume averages used for the other models \citep[e.g.,][]{satyavolu_robustness_2023,anderson_exploring_2026}.

An important point which we will return to later is that none of these models include the extra opacity contributed by the interstellar or circumgalactic gas of the source galaxies. Both Sherwood and Sherwood-Relics use a star-formation scheme optimised for modelling the IGM, whereby any dense ($\rho > 1000 \bar{\rho}$) and cold ($T < 10^5$ K) gas is immediately turned into stars \citep{viel_inferring_2004}. This reduces the computational cost of the simulation while leaving the low-density IGM unaffected. However, as shown in \citet{keating_jwst_2024}, this numerical scheme fails to reproduce the strong DLA profiles  observed in JWST galaxies \citep[e.g.,][]{heintz_extreme_2023}. We address this later in Section \ref{sec:discussion_DLAs}.

For all reference models, we generate mock F070W transmissions by forward-modeling the foreground Lyman-series opacity in the same spirit as \citet{zhu_probing_2023}. At an observed wavelength $\lambda_{\rm obs}$, each Lyman transition $j$ contributes absorption at the foreground redshift
\begin{equation}
    1+z_j = \frac{\lambda_{\rm obs}}{\lambda_j},
\end{equation}
where $\lambda_j$ is the rest-frame wavelength of that transition. The total foreground Lyman-series effective optical depth is then
\begin{equation}
    \tau_{\rm eff}^{\rm LyS}(\lambda_{\rm obs}) =
    \sum_{j={\rm Ly}\alpha}^{j_{\rm max}}
    \tau_{{\rm eff},j}(z_j),
\end{equation}
including only terms with $0<z_j<z_{\rm src}$.
We include all transitions provided by each model, up to the $n=40$ upper level for Sherwood and the first 30 transitions for Sherwood-Relics and FlexRT; truncating the Sherwood series after 30 transitions changes the F070W predictions by less than 0.01\%.

For a source at redshift $z_{\rm src}$, we evaluate this opacity on a rest-frame wavelength grid, $\lambda_{\rm rest}=\lambda_{\rm obs}/(1+z_{\rm src})$, and convert it to transmission,
\begin{equation}
    T_{\rm LyS}(\lambda_{\rm rest},z_{\rm src}) =
    \exp[-\tau_{\rm eff}^{\rm LyS}(\lambda_{\rm rest},z_{\rm src})].
\end{equation}
The model prediction for F070W is then the mean transmission weighted by the photon-counting response to the source continuum,
\begin{equation}
\begin{split}
    T_{\rm F070W}^{\rm model}(z_{\rm src})
    &=
    \frac{
    \int d\lambda_{\rm obs}\,
    R_{\rm F070W}(\lambda_{\rm obs})\,
    \lambda_{\rm obs}^{\alpha-1}\,
    T_{\rm LyS}
    }
    {
    \int d\lambda_{\rm obs}\,R_{\rm F070W}(\lambda_{\rm obs})\,\lambda_{\rm obs}^{\alpha-1}
    }, \\
    T_{\rm LyS}
    &\equiv
    T_{\rm LyS}(\lambda_{\rm obs}/[1+z_{\rm src}],z_{\rm src}) ,
\end{split}
\label{eq:model_f070w}
\end{equation}
where $R_{\rm F070W}$ is the filter throughput and $\alpha$ is the fitted continuum slope of the galaxy, $f_\nu\propto\lambda^{\alpha}$ (Section~\ref{sec:data_jades}), so that the model and the measured $T_{\rm F070W}$ refer to the same band-averaged quantity. We evaluate this prediction for each galaxy at its own redshift and continuum slope and average the predictions within each redshift bin. The integral is evaluated on a rest-frame grid with 0.5~\AA\ spacing, which is converged to better than 1\%.

The galaxy calculation differs from the quasar mean-free-path calculation of \citet{zhu_probing_2023} in two ways. We omit the quasar proximity-zone term, since the background sources are galaxies. We set the transmitted flux below the Lyman limit to zero, because the escaping Lyman-continuum flux of these galaxies is small.

\section{Measurements and Results} \label{sec:results}

\begin{figure*}[!ht]
    \centering
    \includegraphics[width=\textwidth]{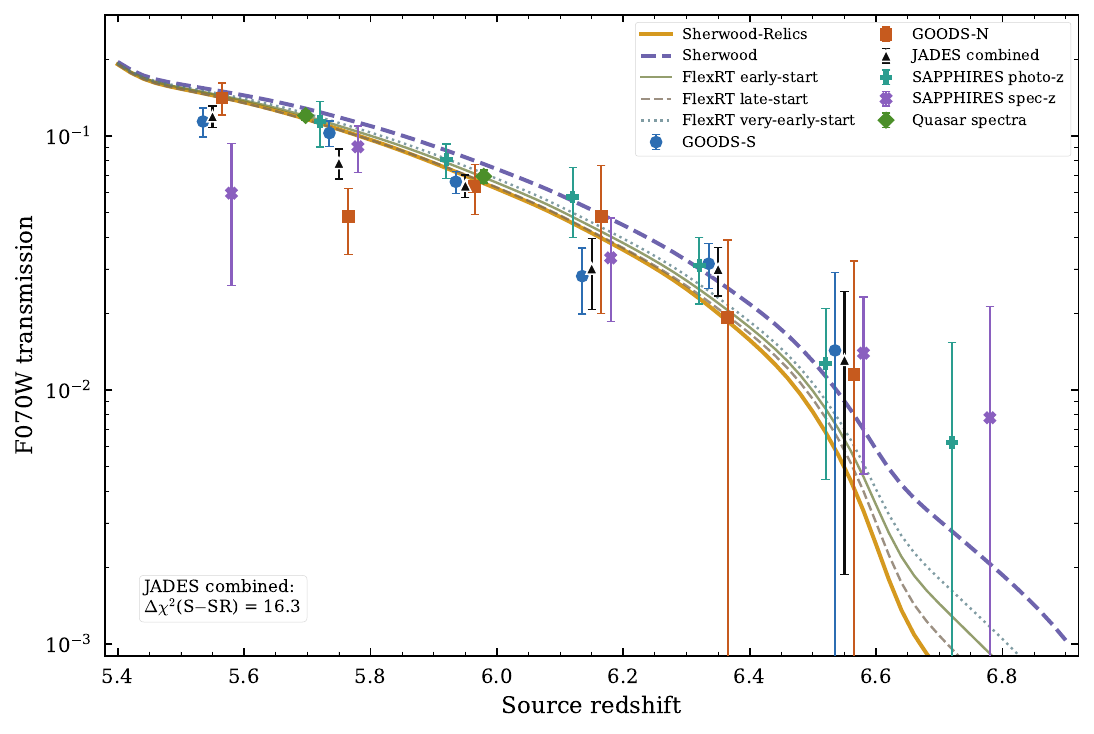}
    \caption{
    Redshift evolution of the F070W transmission measurement.
    Points show GOODS-S, GOODS-N, JADES combined, SAPPHIRES photo-$z$, SAPPHIRES spec-$z$, and filter-weighted quasar measurements; error bars show the bootstrap standard deviations used in the fits.
    Measurements below $9\times10^{-4}$, including all negative values, are omitted from display but retained in the analysis.
    Curves show the Sherwood baseline, the Sherwood-Relics late-reionization model, and three FlexRT reionization histories \citep{cain_chasing_2025,gangolli_correlation_2025}, all forward-modeled through the F070W filter for the median continuum slope of the sample.
    Small horizontal offsets are applied only for visual clarity; table values use the true redshift-bin centers.
    The annotation gives $\Delta\chi^2=\chi^2_{\rm Sherwood}-\chi^2_{\rm Sherwood\mbox{-}Relics}$ for the seven JADES combined bins with bootstrap errors.
    }
    \label{fig:model_comparison}
\end{figure*}

\begin{deluxetable*}{rrrrrr}
\tablecaption{JADES Combined F070W Transmission Measurements \label{tab:jades_binned}}
\tablehead{
\colhead{$z_{\rm bin}$} &
\colhead{$N_{\rm src}$} &
\colhead{$T_{\rm F070W}$} &
\colhead{$T_{16}$} &
\colhead{$T_{84}$} &
\colhead{$\tau_{\rm eff}$}
}
\startdata
5.55 & 39 & 0.1197 & 0.1084 & 0.1303 & 2.12 \\
5.75 & 29 & 0.0784 & 0.0682 & 0.0890 & 2.55 \\
5.95 & 98 & 0.0639 & 0.0575 & 0.0702 & 2.75 \\
6.15 & 25 & 0.0301 & 0.0207 & 0.0401 & 3.50 \\
6.35 & 23 & 0.0299 & 0.0235 & 0.0365 & 3.51 \\
6.55 & 45 & 0.0132 & 0.0023 & 0.0248 & 4.33 \\
6.75 & 20 & $<0.0226$ & \ldots & \ldots & $>3.79$ \\
\enddata
\tablecomments{
$T_{\rm F070W}$ is the bootstrap mean of the $3\sigma$-clipped means of the selected CIRC1 galaxy transmissions.
$T_{16}$ and $T_{84}$ are bootstrap percentiles.
$\tau_{\rm eff}=-\ln T_{\rm F070W}$. For the $z=6.75$ bin, whose mean transmission is negative ($-0.0068$), we list the $2\sigma$ upper limit on $T_{\rm F070W}$ and the corresponding lower limit on $\tau_{\rm eff}$; the model fits use the signed measurement.
The complete field, SAPPHIRES, and quasar-comparison measurements are available in the machine-readable table.
}
\end{deluxetable*}

\subsection{Continuum-normalized F070W Transmission} \label{sec:measurement}\label{sec:continuum}

For each galaxy, we estimate the intrinsic F070W continuum from the eligible redward CIRC1 photometry described in Section~\ref{sec:data_jades}. The fitted power law is averaged over the full F070W response, so that it refers to the same band-averaged quantity as the observed flux; evaluating it instead at the F070W pivot wavelength changes $T_{\rm F070W}$ by less than 0.5\%.

The source-level F070W transmission is defined as
\begin{equation}
    T_{\rm F070W} =
    \frac{F_{\rm F070W,obs}}{F_{\rm F070W,cont}},
\end{equation}
where $F_{\rm F070W,obs}$ is the observed CIRC1 F070W flux and $F_{\rm F070W,cont}$ is the band-averaged continuum prediction. We keep negative source-level values for selected sources to preserve the noise distribution around very low transmission. The fiducial binned value is the bootstrap sigma-clipped mean, with the 16th and 84th bootstrap percentiles used as uncertainties. The unclipped mean differs by at most 0.008 ($0.7\sigma$). Appendix~\ref{app:photoz_robustness} propagates the available photo-$z$/spec-$z$ scatter through this redshift-dependent binning and continuum selection.

To examine the continuum reconstruction, we also fit the UV slope $\beta$ from clean redward JADES photometry. The 279 galaxies have a median $\beta=-2.28$ and 16--84 percentile range $[-2.50,-1.93]$. For each source, we anchor a sampled UV-slope continuum to the nearest clean redward band. For the 227 sources with positive finite transmission estimates in both methods, the resulting transmissions differ from the fiducial estimates by a median $\Delta\log_{10}T_{\rm F070W}=0.004$, with a normalized median absolute deviation (NMAD) of 0.053. Figure~\ref{fig:continuum_robustness} shows the comparison.

One caveat is the intrinsic continuum below Ly$\alpha$, which we extrapolate from the redward bands. For quasars, power-law reconstructions scatter by $\simeq13\%$ over the Ly$\alpha$ forest and $\simeq20\%$ over the Ly$\beta$ forest \citep{bosman_comparison_2021}. As a test, we forward-model the galaxies with BC03, BPASS, and FSPS spectra matched to the observed $\beta$ range (Appendix~\ref{app:continuum_robustness}). These templates fall below the redward power law at 912--1216~\AA; if the galaxies followed them, the recovered transmission would be 20--27\% lower at $z\leq6.35$ and 32--35\% lower at higher redshift. The far-ultraviolet spectra of high-redshift galaxies remain poorly constrained, and future observations and models will help quantify this caveat. Two smaller effects add F070W flux. A background residual, estimated from sources at $z_a>7.7$ whose F070W flux should vanish, could raise $T_{\rm F070W}$ by $\approx0.005$--0.01 in each bin, and intrinsic Ly$\alpha$ emission adds at most $\approx0.005$ in the lowest-redshift bin.

\subsection{Evolution of the F070W Transmission} \label{sec:results_evolution}

Figure~\ref{fig:model_comparison} shows that the combined JADES F070W transmission declines with increasing source redshift, from $T_{\rm F070W}=0.1197$ at $z_{\rm bin}=5.55$ to $0.0132$ at $z_{\rm bin}=6.55$ (Table~\ref{tab:jades_binned}); the $z=6.75$ bin is consistent with zero within its bootstrap interval. GOODS-S and GOODS-N show the same broad evolution and are statistically consistent overall ($\chi^2=11.7$ for seven bins), with the largest difference at $z_{\rm bin}=5.75$ ($3.0\sigma$). The SAPPHIRES samples contain only 2--13 sources per bin and have larger uncertainties; their photo-$z$ measurements agree with JADES within $1.4\sigma$ in every bin, and the spec-$z$ measurements agree except in one three-source bin at $z=6.35$.

We forward-model the two quasar composites of \citet{zhu_probing_2023}, at $z=5.65$ and 5.93, through the same F070W rest-frame window and photometric weighting as the galaxies. They give $T_{\rm F070W}=0.121\pm0.005$ and $0.069\pm0.004$, respectively. Interpolated to the same redshifts, the JADES transmission is $0.82\pm0.10$ and $0.94\pm0.11$ times these values, consistent within $1.9\sigma$.

\subsection{Comparison to Forward-modeled Opacity Models} \label{sec:results_models}

With the power-law continuum, the JADES measurements favor the Sherwood-Relics late-reionization model over the homogeneous-UVB Sherwood baseline. Sherwood predicts higher transmission over much of the measured redshift range, while Sherwood-Relics follows the observed decline more closely. The three FlexRT histories give intermediate predictions at $z_{\rm src}\gtrsim6.3$. Appendix~\ref{app:model_order_transmission} shows the corresponding Lyman-series transmission grids.

We quantify the relative preference using
\begin{equation}
    \Delta\chi^2 =
    \chi^2_{\rm Sherwood}-\chi^2_{\rm Sherwood\mbox{-}Relics},
\end{equation}
evaluated for all seven JADES combined bins in linear transmission, including the negative $z=6.75$ bin, with the bootstrap errors as diagonal uncertainties. We find $\chi^2_{\rm Sherwood\mbox{-}Relics}=18.5$ and $\chi^2_{\rm Sherwood}=34.8$ for seven degrees of freedom, and $\Delta\chi^2=16.3$; the FlexRT histories give $\chi^2=17.8$--24.7. Because the fixed curves leave significant residuals ($\chi^2/{\rm dof}=2.5$--5.0), we use $\Delta\chi^2$ as a relative measure of model preference.

We also evaluate the comparison after adding half of the per-bin GOODS-S/GOODS-N difference in quadrature as an empirical field-scatter term. With this error model, $\chi^2_{\rm Sherwood\mbox{-}Relics}=6.6$, $\chi^2_{\rm Sherwood}=13.4$, and $\Delta\chi^2=6.8$.

\begin{figure*}[!ht]
    \centering
    \includegraphics[width=\textwidth]{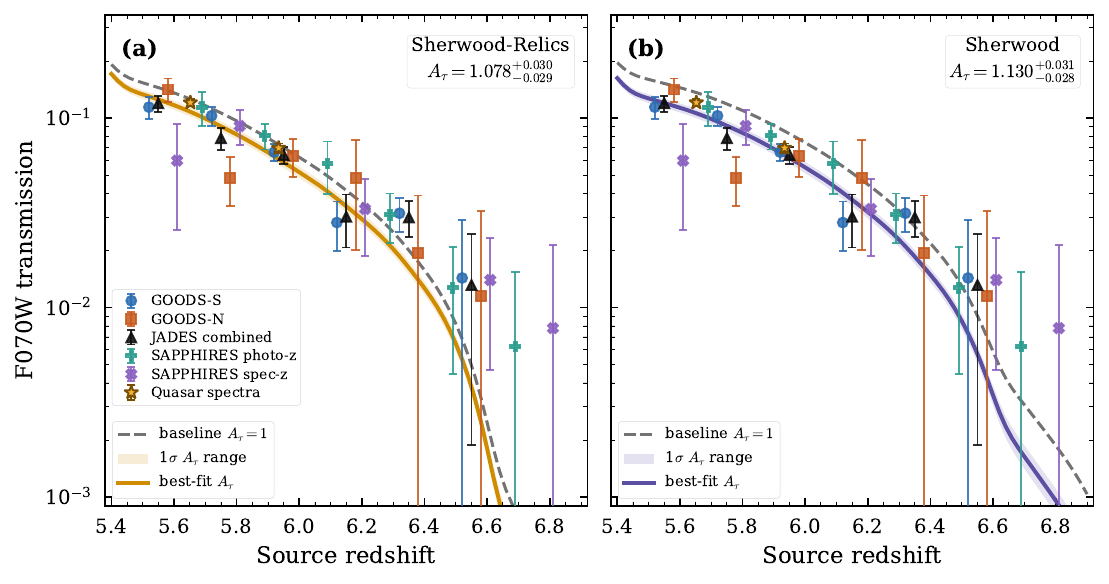}
    \caption{
    Effective opacity-amplitude fits.
    Panels (a) and (b) show the Sherwood-Relics and Sherwood opacity grids, respectively.
    Dashed curves use the baseline $A_{\tau}=1$ model, while solid curves rescale the total Lyman-series optical depth by the best-fit effective amplitude for the JADES combined bins.
    Shaded bands show the 1$\sigma$ uncertainty range from the fitted $A_{\tau}$ values.
    The $A_{\tau}$ fits use all seven JADES combined measurements, shown on the same axis as Figure~\ref{fig:model_comparison}. GOODS-S, GOODS-N, SAPPHIRES, and quasar measurements are shown for comparison; the quasar measurements are not included in the galaxy fits. As in Figure~\ref{fig:model_comparison}, measurements below $9\times10^{-4}$ are omitted from display but retained in the analysis.
    }
    \label{fig:atau_fit}
\end{figure*}

We next fit an effective optical-depth amplitude,
\begin{equation}
    T_{\rm model}(z_{\rm src}; A_{\tau}) =
    \frac{\sum_{i\in\mathcal{B}} w_i \exp[-A_{\tau}\tau_{\rm total}(\lambda_i, z_{\rm src})]}
         {\sum_i w_i},
\end{equation}
where $\mathcal{B}$ contains pixels at rest-frame wavelengths above the source Lyman limit, and $w_i$ includes the F070W throughput, the continuum weighting of Equation~(\ref{eq:model_f070w}), and the wavelength sampling weights. Pixels below the Lyman limit retain their weights in the denominator. Because F070W includes foreground Ly$\alpha$ as well as Ly$\beta$ and higher-order absorption, $A_\tau$ describes the total broadband opacity. For the JADES combined bins, Sherwood-Relics requires $A_{\tau}=1.078^{+0.030}_{-0.029}$ and the Sherwood baseline requires $A_{\tau}=1.130^{+0.031}_{-0.028}$ (Figure~\ref{fig:atau_fit}), with uncertainties from the $\Delta\chi^2=1$ range. Sherwood-Relics thus needs a smaller correction, consistent with its preference in the fixed-model comparison. The fitted $A_{\tau}$ is an empirical amplitude of the total modeled F070W opacity and depends on the assumed continuum, background, and source-associated absorption (Sections~\ref{sec:continuum} and \ref{sec:discussion_DLAs}).

\section{Discussion} \label{sec:discussion}

\begin{figure*}[!ht]
    \centering
    \includegraphics[width=\textwidth]{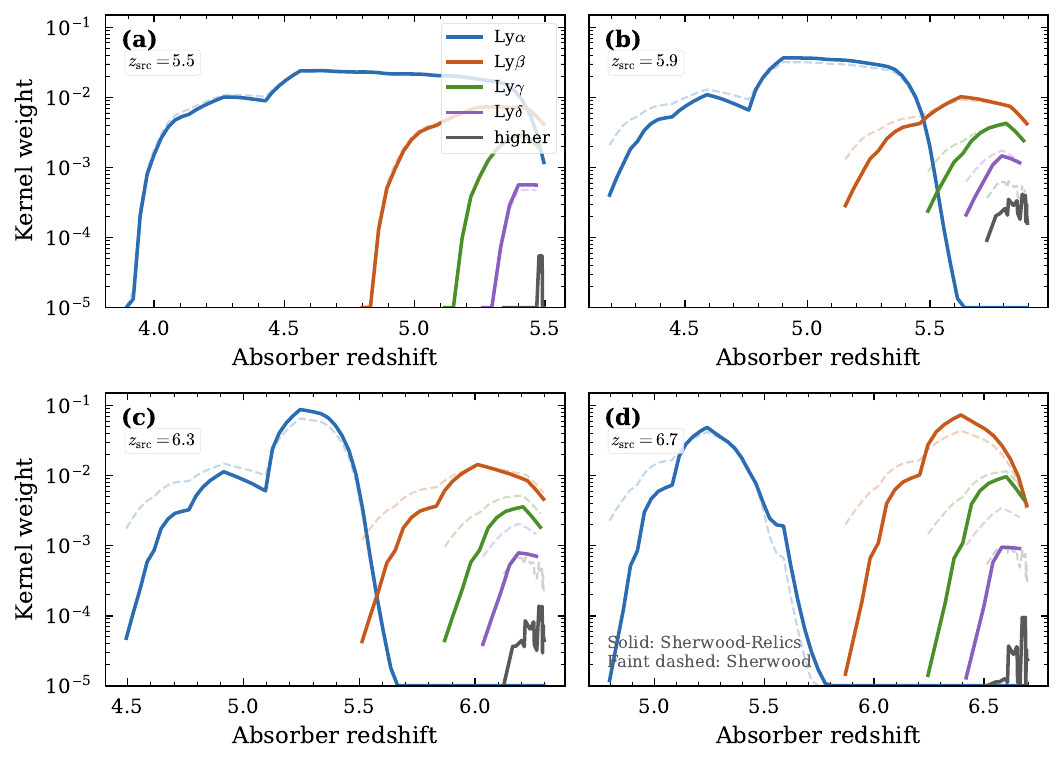}
    \caption{
    Lyman-series response kernels for the F070W measurement.
    Panels (a)--(d) show the normalized sensitivity-kernel weight as a function of absorber redshift for representative source redshifts.
    The colored curves split the contribution into Ly$\alpha$, Ly$\beta$, Ly$\gamma$, Ly$\delta$, and higher-order Lyman-series terms; solid curves show the Sherwood-Relics late-reionization model, and faint dashed curves show the Sherwood \citep{bolton_sherwood_2017} model.
    }
    \label{fig:kernels}
\end{figure*}

\subsection{F070W as a Broadband LyS Opacity Probe} \label{sec:discussion_kernel}

Quasar spectra resolve the LyS forest along individual sightlines, but suitable high-redshift quasars are rare. Deep JWST/NIRCam imaging measures the broadband transmission toward hundreds of galaxies in JADES and SAPPHIRES, reaching a precision of $\sim0.01$ per redshift bin in JADES.

Figure~\ref{fig:kernels} shows which Lyman transitions and absorber redshifts contribute to the F070W transmission. Foreground Ly$\alpha$ is important at lower source redshifts, while Ly$\beta$ contributes much of the higher-order absorption. Ly$\gamma$ and higher transitions become increasingly important toward the Lyman limit. F070W therefore measures the combined LyS opacity across a broad range of foreground redshifts.

\subsection{Ly$\beta$ Comparison} \label{sec:discussion_lybeta}

Because Ly$\beta$ contributes substantially to the higher-order F070W response, we compare the unscaled models with quasar measurements of the observed Ly$\beta$-region opacity. These measurements also contain foreground Ly$\alpha$ absorption at the same observed wavelength, so the model prediction is
\begin{equation}
    \tau_{\rm eff, Ly\beta}^{\rm obs}(z_{\beta}) =
    \tau_{\rm eff, Ly\beta}^{\rm pure}(z_{\beta})
    + \tau_{\rm eff, Ly\alpha}^{\rm fg}(z_{\alpha,{\rm fg}}) ,
    \label{eq:lybeta}
\end{equation}
where $1+z_{\alpha,{\rm fg}}=(1+z_\beta)\lambda_{{\rm Ly}\beta}/\lambda_{{\rm Ly}\alpha}$. Both terms are taken from the same model, and adding the two effective optical depths assumes that the Ly$\beta$ and foreground Ly$\alpha$ absorbers, separated by several hundred comoving Mpc, are uncorrelated. We do not rescale either term by $A_\tau$, because the F070W amplitude describes the total broadband opacity and does not isolate the Ly$\beta$ contribution.

\begin{figure}[!ht]
    \centering
    \includegraphics[width=\columnwidth]{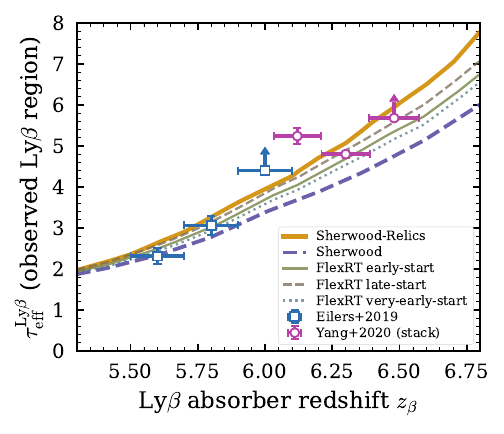}
    \caption{
    Observed Ly$\beta$-region effective optical depth compared with quasar Ly$\beta$ measurements.
    The unscaled ($A_\tau=1$) model curves include Ly$\beta$ opacity at $z_\beta$ plus foreground Ly$\alpha$ opacity at the same observed wavelength (Equation~\ref{eq:lybeta}).
    Quasar points show the binned values of \citet{eilers_anomaly_2019} and the stacked values of \citet{yang_measurements_2020}, both measured in the observed Ly$\beta$ region; the Eilers point at $z_\beta=6.0$ and the Yang point at $z_\beta=6.48$ are $2\sigma$ lower limits.
    Horizontal bars show the published redshift bins (for \citealt{yang_measurements_2020}, the $50\,h^{-1}$~cMpc bins converted to redshift).
    }
    \label{fig:high_order_decomposition}
\end{figure}

Figure~\ref{fig:high_order_decomposition} compares these predictions with quasar Ly$\beta$-region measurements. At $z_\beta\lesssim5.8$ all models agree with \citet{eilers_anomaly_2019} within $\approx2\sigma$. At $z_\beta\simeq6.0$--6.1 the measured opacities exceed all unscaled models. At $z_\beta\simeq6.3$, the FlexRT late-start and early-start histories lie within $1.3\sigma$ of the stacked measurement of \citet{yang_measurements_2020}, whereas Sherwood-Relics is more opaque and the FlexRT very-early-start history and the Sherwood baseline are more transparent. At $z_\beta\simeq6.48$, only Sherwood-Relics exceeds the \citet{yang_measurements_2020} lower limit. Hashemi et al.\ (in preparation) also find that a late Sherwood-Relics history better reproduces quasar composite spectra spanning both LyS and Lyman-continuum absorption. Quasar radiation can ionize gas close to the source; for galaxy sightlines, neutral gas in the ISM and CGM may contribute appreciably to the LyS opacity (Section~\ref{sec:discussion_DLAs}).

\begin{figure*}[!ht]
    \centering
    \includegraphics[width=\textwidth]{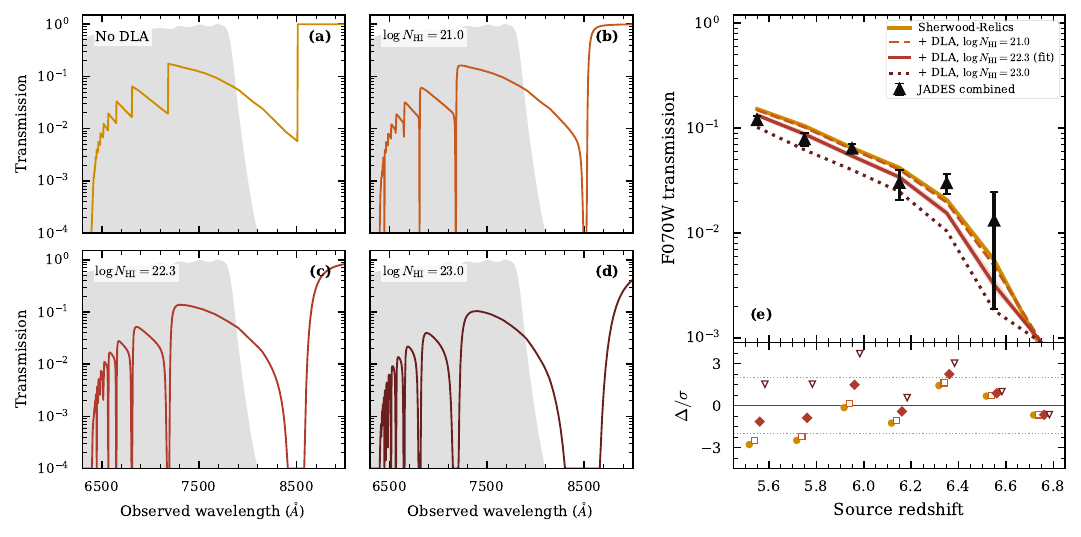}
    \caption{
    Source-associated H\,\textsc{i} absorption.
    Panels (a)--(d) show the Sherwood-Relics transmission toward a source at $z=6$ without a DLA and with a source-associated DLA of $\log_{10}(N_{\rm HI}/{\rm cm}^{-2})=21$, 22.3 (best fit), and 23; the shaded region shows the normalized F070W throughput.
    In panel (e), triangles show the JADES combined measurements with bootstrap errors, the solid orange curve shows the Sherwood-Relics prediction, and the red and brown curves add the DLA at the redshift of every galaxy (full covering); the red band shows the $1\sigma$ range of the fitted column.
    All predictions in panel (e) are evaluated for each galaxy and averaged within each redshift bin.
    The lower part of panel (e) shows the residuals, (data$-$model)/$\sigma$, for each curve (filled symbols: Sherwood-Relics without a DLA and with the best-fit DLA); dotted lines mark $\pm2\sigma$.
    }
    \label{fig:dla_test}
\end{figure*}

\subsection{Late Reionization and Additional Intergalactic Opacity} \label{sec:discussion_implications}

Our F070W measurements trace the rapid decline in broadband LyS transmission at $z\simeq5.5$--6.8. Assuming a power-law intrinsic continuum, late-reionization models reproduce this evolution better than the homogeneous-UVB Sherwood baseline. The global opacity fits give $A_\tau=1.078$ for Sherwood-Relics and 1.130 for Sherwood (Section~\ref{sec:results_models}), so both models underpredict the observed opacity. The additional absorption can arise in the IGM or near the galaxies.

The excess is measured in only two fields. Each JADES field spans $\approx34$--37 comoving Mpc on a side at $z\simeq6$, so the bootstrap errors do not capture cosmic variance between fields. Both fields show an excess of sources near $z_a\simeq6$ with similar fractional amplitude. Measurements in more independent fields, with spectroscopic redshifts and constraints on galaxy far-ultraviolet continua, will help distinguish global opacity evolution from field-to-field variation. Future Roman grism surveys could provide the needed area and redshift information.

Additional intergalactic opacity could arise from residual neutral islands near the end of reionization. Damping wings associated with long dark gaps provide evidence for substantially neutral gas below $z=6$ \citep[e.g.,][]{zhu_damping_2024,becker_damping_2024,spina_damping_2024,gnedin_damping_2025}. Small-scale self-shielded structures can also increase the opacity and shorten the ionizing mean free path, consistent with an absorption-dominated end to reionization and recent constraints on IGM clumping \citep{park_hydrodynamic_2016,chardin_self-shielding_2018,daloisio_hydrodynamic_2020,nasir_hydrodynamic_2021,davies_predicament_2021,austin_resolving_2025,davies_predicament_2026}.
Temperature and UVB fluctuations within the ionized IGM can also contribute \citep{mcquinn_signatures_2011,daloisio_large_2015,daloisio_large_2018,gaikwad_probing_2020}. Joint comparisons of the Ly$\alpha$ forest, broadband LyS opacity, and the ionizing mean free path, combined with higher-resolution radiation-hydrodynamic simulations, can separate the contributions of large-scale neutral regions, small-scale absorbers, and thermal structure.

\subsection{The Contribution of Source-Associated Neutral Hydrogen} \label{sec:discussion_DLAs}

Source-associated neutral gas is another possible contributor. Post-reionization, galaxies are known to host significant amounts of neutral gas, e.g. at $z\sim3$ \citet{reddy_connection_2016} find evidence for partial covering of Lyman-break galaxies with $\log_{10}(N_{\rm HI}/{\rm cm}^{-2}) \approx 21$. Reionization-era galaxies appear to frequently host neutral gas of similar H\,\textsc{i} column densities \citep{mason_constraints_2026} or perhaps even stronger absorption systems \citep[e.g.,][]{heintz_extreme_2023,hainline_searching_2024,asada_improving_2025,heintz_dense_2026,pollock_characterising_2026}. These galaxy DLAs are also found in cosmological simulations \citep{gelli_neutral_2025,mallik_origin_2026,steen_galaxy_2026}. In spectroscopic forest analyses conducted with galaxies, it is possible to mask wavelengths close to the source \citep[e.g.,][]{meyer_probing_2025}, reducing sensitivity to proximate absorption. As described in Section \ref{sec:data_models}, this source-associated absorption is not included in the reference IGM models.

To test whether absorbers beyond the IGM can account for the additional opacity, we add a DLA following the approximation to the Voigt profile presented in \citet{tepper-garcia_voigt_2006} at each galaxy redshift in the Lyman-series transmission measured in the Sherwood-Relics model. We test two illustrative cases: $\log_{10}(N_{\rm HI}/{\rm cm}^{-2})=21$ and 23. The lower column is typical of local H\,\textsc{i} absorption inferred for high-redshift galaxies \citep{mason_constraints_2026} and the higher column is an extreme case, exceeding any of the measurements from \citet{pollock_characterising_2026}. We assume full covering \citep[consistent with the Lyman-break galaxy results][]{reddy_connection_2016} and include the damping wings from each Lyman-series transition. For each galaxy, we integrate the absorbed spectrum through F070W and refit the continuum to the redward bands as in the main analysis (Figure~\ref{fig:dla_test}).

At $z\leq6.35$, the two cases lower the predicted transmission by 2--6\% and 33--50\%, respectively, compared with the 13--19\% reduction implied by the best-fit $A_\tau$. For all seven bins, the baseline model gives $\chi^2=18.5$, compared with 15.8 and 29.3 for the two DLA columns. With partial covering, the 23 case fits with a covering fraction of $\approx0.4$ (0.27--0.54; $\chi^2=9.5$). These combinations illustrate the source-associated absorption that could account for the residual opacity. We note that at column densities as high as $\log_{10}(N_{\rm HI}/{\rm cm}^{-2})= 23$, the DLA profile we have used begins to lose accuracy, as the asymmetry of the red and blue wings becomes more apparent \citep{lee_quantum_2020} and which is missing in our treatment.

Beyond these fixed cases, we then fit the column density continuously over $\log_{10}(N_{\rm HI}/{\rm cm}^{-2})=20.5$--23.5. Assuming the Sherwood-Relics IGM transmission and full DLA covering, the broadband LyS measurements favor an effective column density of $\log_{10}(N_{\rm HI}/{\rm cm}^{-2})=22.3^{+0.2}_{-0.3}$, with $\chi^2=10.7$ for seven bins compared with 18.5 without a DLA (Figure~\ref{fig:dla_test}e). Because the transmission depends nonlinearly on column density, this single full-covering column is a characteristic value rather than the mean of the underlying column-density distribution. At this column, the \citet{tepper-garcia_voigt_2006} approximation reproduces the exact Voigt profile, and an approximate quantum-mechanical correction to the Ly$\alpha$ wings \citep{lee_quantum_2020} raises the fitted column by only 0.06~dex. With the empirical field-scatter term, this constraint becomes an upper limit, $\log_{10}(N_{\rm HI}/{\rm cm}^{-2})<22.6$ ($2\sigma$). The inferred column also depends on the intrinsic continuum and on the adopted IGM transmission, including the treatment of dense gas and foreground Lyman-series absorption discussed in Appendix~\ref{app:model_order_transmission}. For the homogeneous-UVB Sherwood model, the same fit gives 22.65 ($\chi^2=6.8$), so source-associated and intergalactic absorption remain degenerate, as in damping-wing analyses of individual spectra \citep[e.g.,][]{davies_possible_2023,chen_characteristic_2023,keating_origin_2024}. The late-reionization preference in Section~\ref{sec:results_models} thus holds when source-associated absorption is small. Dense H\,\textsc{i} near galaxies would also suppress their Ly$\alpha$ emission and may contribute to the low Ly$\alpha$ visibility of galaxies in overdense regions during reionization \citep{zhu_low_2026}.

This demonstrates a novel method for measuring the average H\,\textsc{i} column density across a large sample of galaxies, complementary to directly fitting damping wing profile to spectra. Building on this, performing the same test in galaxies binned by redshift or by their properties such as UV magnitude, would reveal whether this absorption varies with time and/or galaxy type. This would help to provide informative priors on source-associated absorption for studies of IGM damping wings to measure the timing of reionization \citep{mason_constraints_2026, umeda_jwst_2023, umeda_probing_2026}. The caveat, however, is that it assumes the mean flux in the IGM simulations is correct. Future work refining the reionization simulations with additional spectroscopic constraints \citep[e.g.,][]{meyer_probing_2025} should provide the data to improve this in the near future.

\section{Conclusions} \label{sec:conclusions}

We measure broadband LyS transmission toward 279 JADES galaxies at $5.45<z<6.85$ with deep JWST/NIRCam F070W imaging. The mean transmission falls from $0.120$ at $z=5.55$ to $0.013$ at $z=6.55$ and is consistent with zero at $z=6.75$. Assuming a power-law intrinsic continuum, we find two main results.
\begin{enumerate}
    \item With no source-associated DLA, the fixed-model comparison favors late reionization. The Sherwood-Relics model fits better than the homogeneous-UVB Sherwood baseline ($\Delta\chi^2=16.3$, or 6.8 with the empirical field-scatter term) and needs a smaller opacity rescaling, $A_\tau=1.078^{+0.030}_{-0.029}$ compared with $1.130^{+0.031}_{-0.028}$ for Sherwood.
    \item Source-associated H\,\textsc{i} can account for the remaining opacity. Assuming the Sherwood-Relics IGM and full DLA covering, we infer an effective column density of $\log_{10}(N_{\rm HI}/{\rm cm}^{-2})=22.3^{+0.2}_{-0.3}$.
\end{enumerate}
The two results are linked, because the inferred column depends on the assumed IGM transmission: a homogeneous-UVB IGM requires a larger column. Better constraints on the intrinsic far-ultraviolet continua of these galaxies and on the IGM transmission will sharpen both results. Deep broadband photometry thus extends LyS transmission measurements to hundreds of galaxy sightlines near the end of reionization and provides a new statistical constraint on the neutral gas in reionization-era galaxies.

\section*{Acknowledgments}

We thank Jakob Helton for helpful discussions on photometric-redshift uncertainties, and Rohan Naidu for valuable suggestions on the presentation of Figure~\ref{fig:dla_test}. We acknowledge support from the NIRCam Science Team contract to the University of Arizona, NAS5-02105. Y.Z. is also supported by JWST Program \#6434. G.D.B. and S.H. were supported by JWST Program \#4092. Support for programs \#6434 and \#4092 was provided by NASA through a grant from the Space Telescope Science Institute, which is operated by the Association of Universities for Research in Astronomy, Inc., under NASA contract NAS 5-03127.

A.J.B.\ acknowledges funding from the ``FirstGalaxies'' Advanced Grant from the European Research Council under the European Union's Horizon 2020 research and innovation programme (grant agreement No. 789056).

L.K. acknowledges the support of a Royal Society University Research Fellowship (grant number URF$\backslash$R1$\backslash$251793).

C.C. acknowledges support from the Beus Center for Cosmic Foundations at Arizona State University.  

G.K.\ acknowledges support by the Department of Atomic Energy, Government of India, under Project Identification Number RTI-4012 and RTI-4013.

This work is based on observations made with the NASA/ESA/CSA James Webb Space Telescope. The data were obtained from the Mikulski Archive for Space Telescopes at the Space Telescope Science Institute, which is operated by the Association of Universities for Research in Astronomy, Inc., under NASA contract NAS 5-03127 for JWST. The JWST data used in this paper can be found in MAST: \dataset[https://doi.org/10.17909/8tdj-8n28]{https://doi.org/10.17909/8tdj-8n28}.

JADES DR5 includes NIRCam data from JWST programs 1176, 1180, 1181, 1210, 1264, 1283, 1286, 1287, 1895, 1963, 2079, 2198, 2514, 2516, 2674, 3215, 3577, 3990, 4540, 4762, 5398, 5997, 6434, 6511, and 6541. JADES DR5 includes MIRI data from JWST programs 1180, 1181, and 1207. The authors acknowledge the teams of programs 1895, 1963, 2079, 2514, 3215, 3577, 3990, 6434, and 6541 for developing their observing program with a zero-exclusive-access period.

We respectfully acknowledge the University of Arizona is on the land and territories of Indigenous peoples. Today, Arizona is home to 22 federally recognized tribes, with Tucson being home to the O'odham and the Yaqui. The university strives to build sustainable relationships with sovereign Native Nations and Indigenous communities through education offerings, partnerships, and community service.

The authors acknowledge the use of ChatGPT and Codex for language editing and code development \citep{openai_chatgpt_2026}. All LLM-assisted outputs were reviewed by the authors, who take full responsibility for the analyses and conclusions.

\facilities{JWST, MAST}

\appendix

\section{Model Lyman-series transmission}
\label{app:model_order_transmission}

\begin{figure*}
    \centering
    \includegraphics[width=\textwidth]{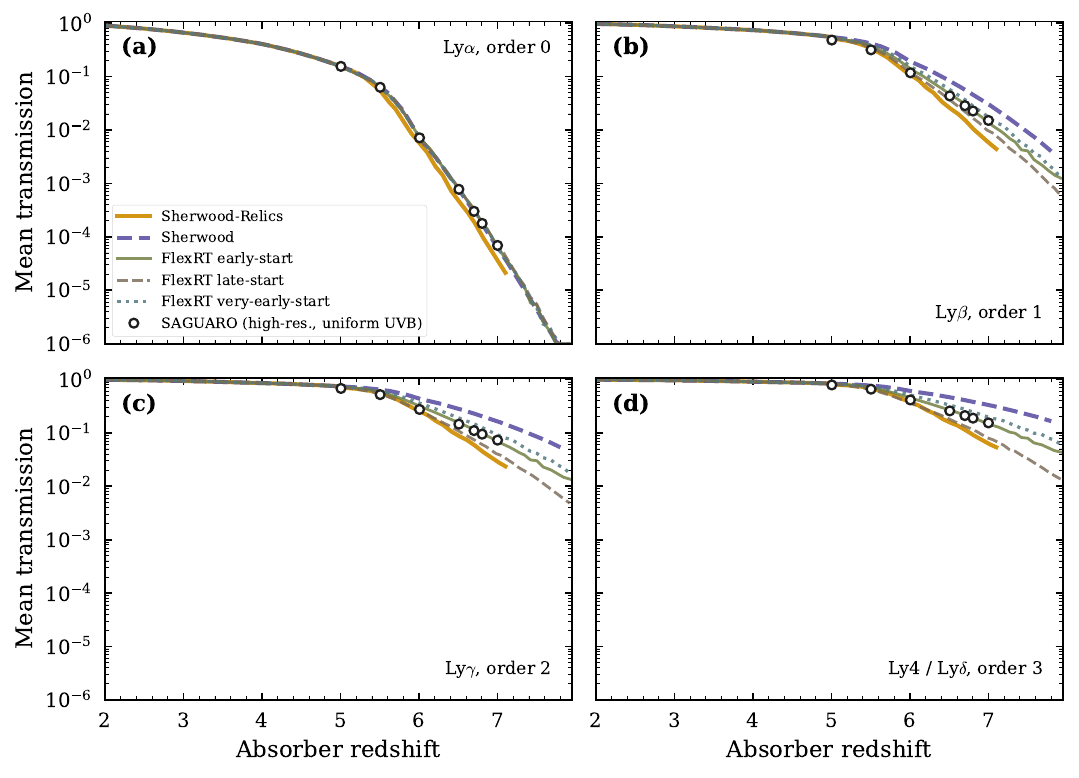}
    \caption{
    Mean transmission as a function of absorber redshift for Ly$\alpha$, Ly$\beta$, Ly$\gamma$, and Ly4/Ly$\delta$.
    Curves show Sherwood, Sherwood-Relics, and the three FlexRT reionization histories, with the Sherwood grid below $z_{\rm abs}=5$ for all models.
    The same model grids are used in the F070W forward modeling shown in Figure~\ref{fig:model_comparison}.
    Open circles show the high-resolution, uniform-UVB SAGUARO calculation at its snapshot redshifts ($z_{\rm abs}\geq5$), for comparison only.
    }
    \label{fig:model_order_transmission}
\end{figure*}

Figure~\ref{fig:model_order_transmission} compares the mean transmission as a function of absorber redshift for the first four Lyman-series orders in the Sherwood, Sherwood-Relics, and FlexRT model grids, as used in the forward modeling. Below $z_{\rm abs}=5$ all curves follow the common Sherwood grid by construction. Above it, the models have similar Ly$\alpha$ transmissions over part of the redshift range; the FlexRT Ly$\alpha$ opacity in ionized gas was rescaled to the \citet{kulkarni_large_2019} model (Section~\ref{sec:data_models}). Their higher-order transmissions differ at $z_{\rm abs}\gtrsim5$. The quick-Ly$\alpha$ approximation used in Sherwood and Sherwood-Relics \citep{viel_inferring_2004} also removes dense, self-shielded gas along intervening sightlines. The common Sherwood foreground at $z_{\rm abs}<5$ is calibrated to the observed mean Ly$\alpha$ transmission, but matching the mean Ly$\alpha$ transmission does not guarantee an accurate H\,\textsc{i} optical-depth distribution or higher-order Lyman-series opacity. Near the end of reionization, $5\lesssim z_{\rm abs}\lesssim5.3$, where each model uses its own grid, the removed gas may also affect the Ly$\alpha$ absorption itself. We also compare with a high-resolution, $2\,h^{-1}$~cMpc SAGUARO simulation with a uniform UVB \citep{cain_introducing_2026}.
At similar mean Ly$\alpha$ transmission, this calculation predicts stronger Ly$\beta$ and Ly$\gamma$ absorption at $z_{\rm abs}\simeq5$--5.5. The difference suggests that small-scale gas structure and thermal conditions can affect the broadband LyS opacity \citep[see also][]{doughty_convergence_2023}.

\section{Survey Footprints} \label{app:footprints}

\begin{figure*}[!ht]
    \centering
    \includegraphics[width=\textwidth]{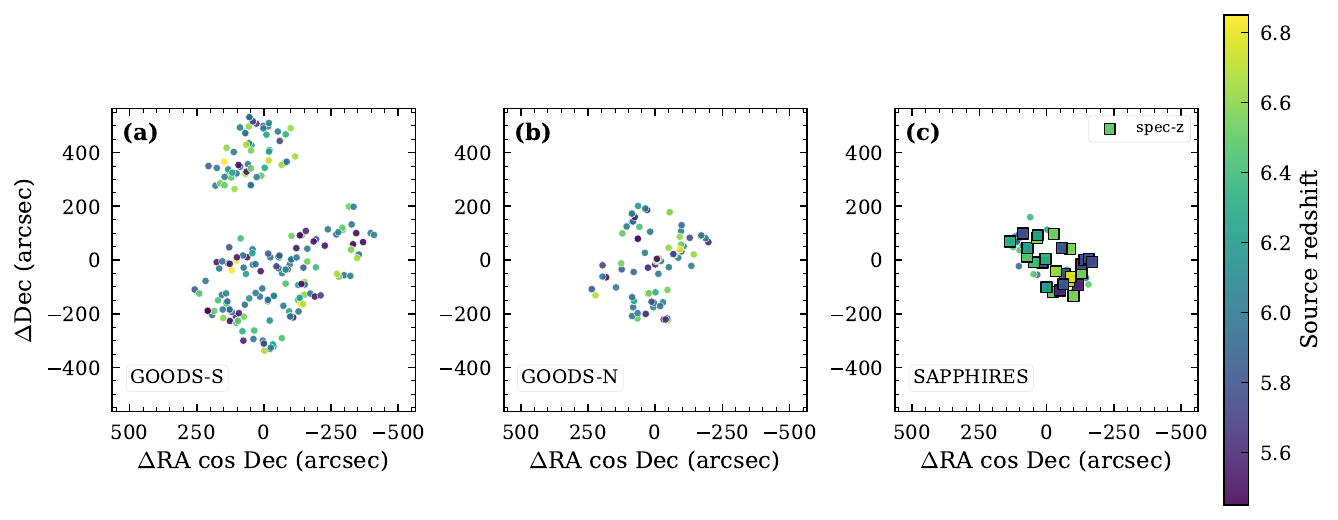}
    \caption{
        Sky distribution of the F070W source samples. Panels (a), (b), and (c) show GOODS-S, GOODS-N, and SAPPHIRES EDR, respectively, with points colored by source redshift. The SAPPHIRES panel shows the photo-$z$ sample and overlays the spec-$z$ sample with square markers. All panels use the same angular scale to compare field size and source sampling.
    }
    \label{fig:field_footprints}
\end{figure*}

Figure~\ref{fig:field_footprints} shows the sky distribution of the F070W-selected sources in GOODS-S, GOODS-N, and SAPPHIRES. The three panels are plotted on the same angular scale to show the relative field sizes and source sampling. 

\section{Photometric-redshift Uncertainty} \label{app:photoz_robustness}

Photometric-redshift errors affect both the source redshift bins and the set of filters allowed in the redward continuum fit. We estimate this effect using the available photo-$z$/spec-$z$ overlap from SAPPHIRES and propagate the empirical redshift residuals through the F070W measurement.

\begin{figure*}[!ht]
    \centering
    \includegraphics[width=\textwidth]{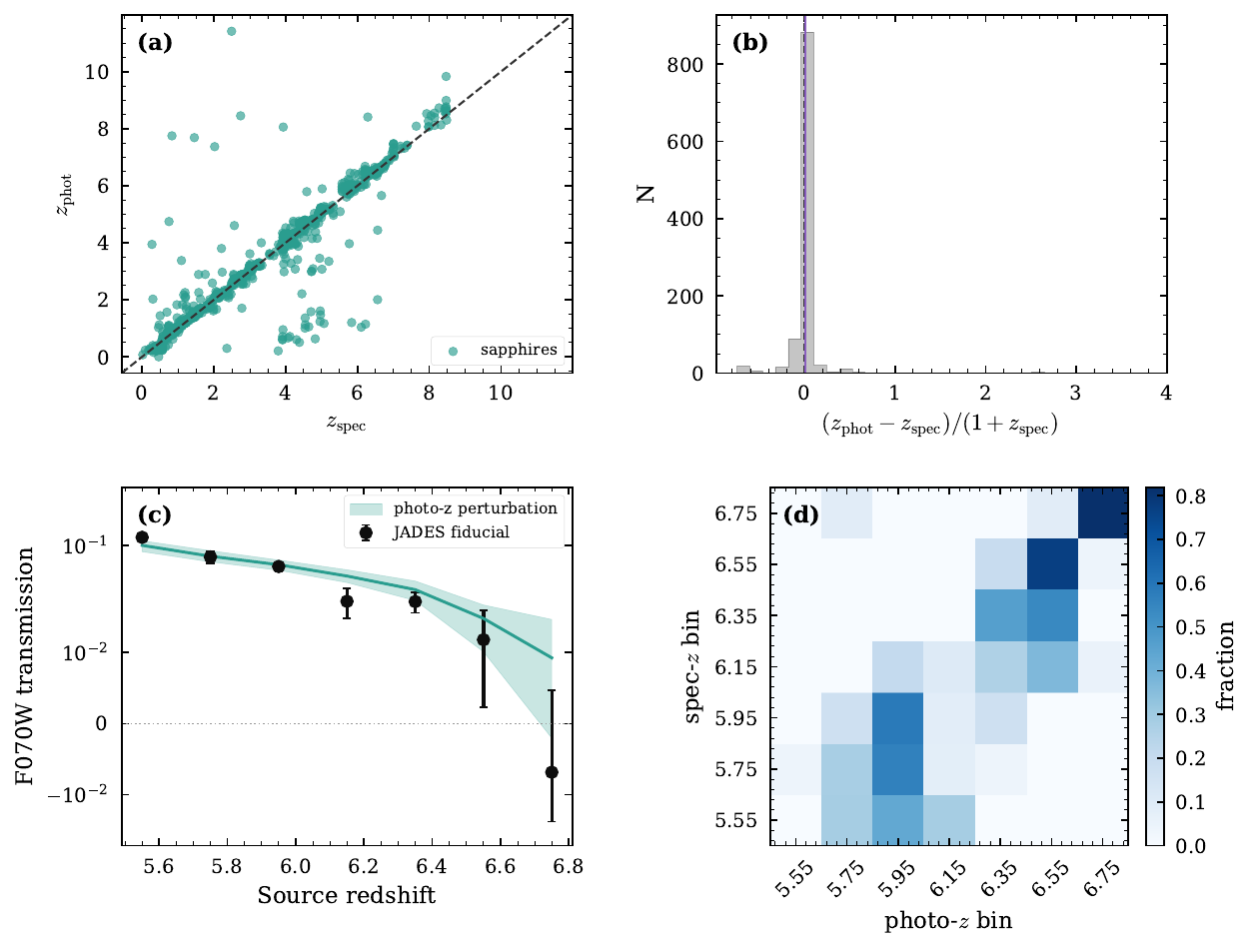}
    \caption{
    Photometric-redshift uncertainty test.
    Panel (a) compares photo-$z$ and spec-$z$ values for the matched sample.
    Panel (b) shows the distribution of $\Delta z/(1+z_{\rm spec})$.
    Panel (c) compares the fiducial JADES combined F070W transmission (Table~\ref{tab:jades_binned}) with the envelope obtained after perturbing source redshifts by the empirical residual distribution; the vertical axis is logarithmic above 0.01 and linear around zero.
    Panel (d) shows the redshift-bin migration matrix for sources in the F070W science range.
    }
    \label{fig:photoz_specz_robustness}
\end{figure*} 

The 1060 SAPPHIRES galaxies with both photometric and spectroscopic redshifts have a median $\Delta z/(1+z_{\rm spec})=0.0079$, a scatter $\sigma_{\rm NMAD}=0.027$ (1.48 times the median absolute deviation of $\Delta z/(1+z_{\rm spec})$), and outlier fractions of 0.216 and 0.112 for $|\Delta z|/(1+z_{\rm spec})>0.05$ and $>0.10$, respectively. After perturbing the JADES redshifts by this empirical residual distribution and recomputing the continuum selection and binning, the declining F070W transmission remains present, although redshift scatter raises the transmission in the bins at $z\geq6.15$ by mixing in lower-redshift sources.

\section{Aperture Dependence} \label{app:aperture_robustness}

To test whether the fiducial CIRC1 aperture affects the result, we repeat the full measurement with fluxes from larger apertures (CIRC2 and CIRC3), using the same aperture for F070W and all redward continuum bands within each run. Because the flux cuts are applied to the larger-aperture fluxes, the selected samples differ: 394 and 652 sources pass the cuts, compared with 279 for CIRC1. For positive JADES combined bins, the median CIRC2/CIRC1 and CIRC3/CIRC1 ratios are 1.03 and 1.36, respectively, and therefore combine aperture and sample differences. In all apertures the F070W transmission declines with redshift and the highest-redshift bins remain near zero; larger apertures mainly shift the normalization.

\begin{figure*}[!ht]
    \centering
    \includegraphics[width=\textwidth]{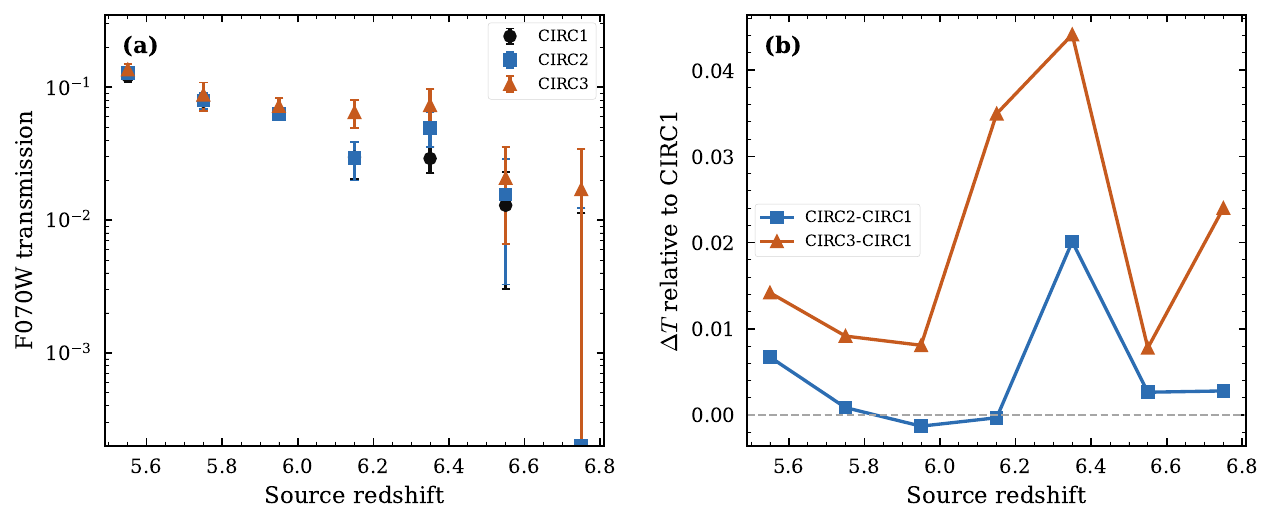}
    \caption{
    Aperture-dependence test.
    Panel (a) compares the JADES combined F070W transmission measured with CIRC1, CIRC2, and CIRC3.
    Panel (b) shows the CIRC2--CIRC1 and CIRC3--CIRC1 transmission differences.
    }
    \label{fig:aperture_robustness} 
\end{figure*}

\section{Continuum Reconstruction} \label{app:continuum_robustness}

\begin{figure*}[!ht]
    \centering
    \includegraphics[width=\textwidth]{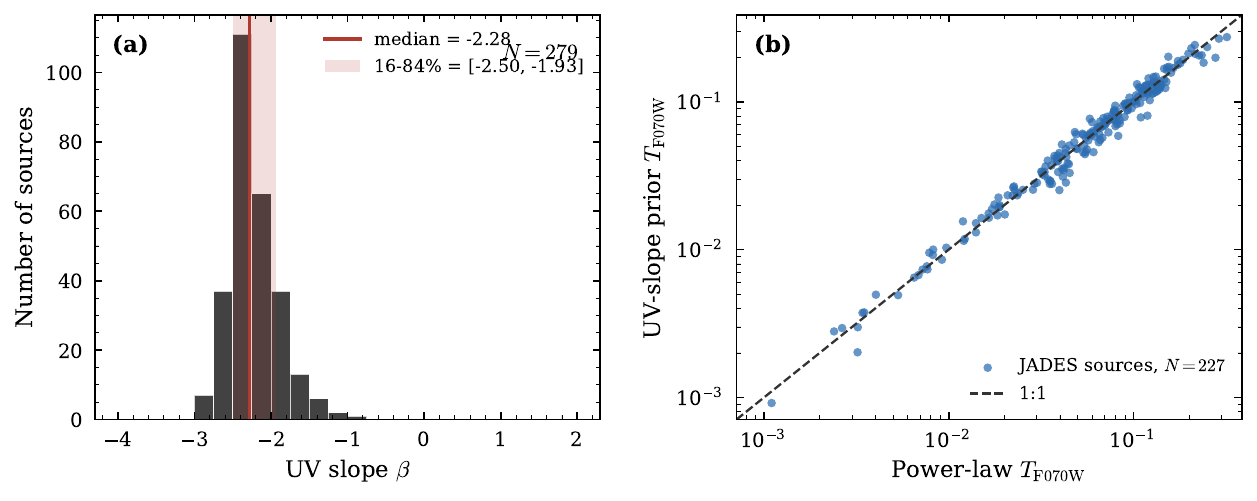} 
    \caption{
    Continuum-reconstruction comparison.
    Panel (a) shows the empirical UV-slope distribution measured from clean redward JADES photometry.
    Panel (b) compares source-level $T_{\rm F070W}$ inferred from the fiducial redward power-law continuum with $T_{\rm F070W}$ inferred from the UV-slope-prior continuum.
    The dashed line shows the one-to-one relation.
    Only sources with positive finite transmission estimates in both methods are shown.
    }
    \label{fig:continuum_robustness}
\end{figure*}

The F070W transmission relies on extrapolating the redward continuum, which we test in two ways. First, Figure~\ref{fig:continuum_robustness} shows the empirical JADES UV-slope distribution and compares source-level F070W transmissions inferred from the fiducial redward power-law and UV-slope-prior continua.

\begin{figure}[!ht]
    \centering
    \includegraphics[width=\columnwidth]{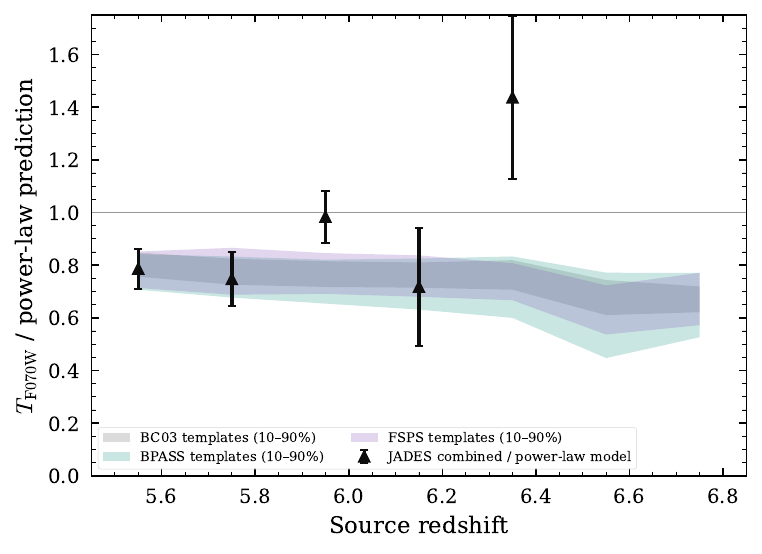}
    \caption{
    Intrinsic-continuum diagnostic.
    Bands show the ratio of the measured to the power-law-predicted $T_{\rm F070W}$ (10th--90th percentiles over templates) when the galaxies are forward-modeled with BC03, BPASS, or FSPS spectra in the observed $\beta$ range and the Sherwood-Relics IGM.
    Triangles show the JADES combined measurements divided by the power-law prediction for bins where the predicted transmission exceeds 0.01.
    }
    \label{fig:continuum_shape}
\end{figure}

Second, Figure~\ref{fig:continuum_shape} tests the assumption that the intrinsic continuum below Ly$\alpha$ follows the redward power law. For each galaxy, we forward-model the photometry with intrinsic stellar-population spectra from BC03 \citep[][through \textsc{Bagpipes}; \citealt{carnall_inferring_2018}]{bruzual_stellar_2003}, BPASS binary populations \citep{eldridge_binary_2017,stanway_re-evaluating_2018}, and FSPS with MIST isochrones \citep{conroy_propagation_2009,choi_mesa_2016}, restricted to $-2.50\leq\beta\leq-1.90$, combined with the Sherwood-Relics IGM, and measure $T_{\rm F070W}$ exactly as for the data.
All templates fall below the redward power law at 912--1216~\AA, so the transmission recovered with our power-law procedure is 20--27\% lower than for a power-law continuum with the same IGM at $z\leq6.35$, and about one third lower at higher redshift. Using one such template across the sample gives acceptable fits for many of the template choices and reverses the model ranking for most of the tested templates (92\%; median $\Delta\chi^2\approx-5$). Thus, the model preference in Section~\ref{sec:results_models} depends on the intrinsic continuum shape.

\end{document}